%% file: EUSIPCO_ver.tex
\documentclass[conference]{IEEEtran}
\IEEEoverridecommandlockouts
\usepackage{cite}
\usepackage{amsmath,amssymb,amsfonts}
\usepackage{algorithmic}
\usepackage{graphicx}
\usepackage{textcomp}
\usepackage{xcolor}
\usepackage{subfig}
\usepackage{units}
\usepackage{mathtools}

\DeclarePairedDelimiter{\ceil}{\lceil}{\rceil}

\usepackage[hidelinks]{hyperref}
\usepackage{pgfplots}
\usepackage{tikz}
\usepackage[T1]{fontenc}
\usepackage[utf8]{inputenc}
\usepackage{pgfplots}
\usepackage{grffile}
\pgfplotsset{compat=newest}
\usetikzlibrary{plotmarks}
\usetikzlibrary{arrows.meta}
\usepgfplotslibrary{patchplots}

\usepackage{url}
\usepackage{adjustbox}
\usepackage[labelsep=period]{caption}
\usepackage{booktabs,array}
\usepackage{multirow}
\usepackage{balance}

\definecolor{red2}{rgb}{1,0.141176471,0.18823529411}
\definecolor{blue2}{rgb}{0.0156862745,0.51372549019,0.96078431372}
\definecolor{purple2}{rgb}{0.733,0.5255,0.9882}
\definecolor{green2}{rgb}{0.0118,0.8549,0.7765}

\newcommand{\myelipsegreen}[1][0.1pt]
{   \begin{tikzpicture}[overlay]
	\draw [line width=0.5mm,green2](0, 0.1) ellipse (1mm and 1mm);
	\end{tikzpicture}
}

\newcommand{\myarrowred}[1][0.1pt]
{   \begin{tikzpicture}[overlay]
	\draw [->,>=stealth,line width=0.5mm,red2] (0, 0.3) -- (0, -0.1);
	\end{tikzpicture}
}

\newcommand{\myarrowblue}[1][0.1pt]
{   \begin{tikzpicture}[overlay]
	\draw [->,>=stealth,line width=0.5mm,blue2] (0, 0.3) -- (0, -0.1);
	\end{tikzpicture}
}

\begin{document}

\title{AeGAN: Time-Frequency Speech Denoising\\ via Generative Adversarial Networks
}

\author{\IEEEauthorblockN{Sherif Abdulatif$\,^{\star}$, Karim Armanious$^{\star}$, Karim Guirguis, Jayasankar T. Sajeev, Bin Yang\thanks{$^{\star}$These authors contributed to this work equally.}}
\IEEEauthorblockA{University~of~Stuttgart,~Institute~of~Signal~Processing~and~System~Theory,~Stuttgart,~Germany}}

\maketitle

\begin{abstract}
Automatic speech recognition (ASR) systems are of vital importance nowadays in commonplace tasks such as speech-to-text processing and language translation. This created the need for an ASR system that can operate in realistic crowded environments. Thus, speech enhancement is a valuable building block in ASR systems and other applications such as hearing aids, smartphones and teleconferencing systems. In this paper, a generative adversarial network (GAN) based framework is investigated for the task of speech enhancement, more specifically speech denoising of audio tracks. A new architecture based on CasNet generator and an additional feature-based loss are incorporated to get realistically denoised speech phonetics. Finally, the proposed framework is shown to outperform other learning and traditional model-based speech enhancement approaches.
\end{abstract}


\section{Introduction}
\label{sec:intro}

In a noisy environment, a typical speech signal is perceived as a mixture between clean speech and an intrusive background noise. Accordingly, speech denoising is interpreted as a source separation problem, where the goal is to separate the desired audio signal from the intrusive noise. The background noise type and the signal-to-noise ratio (SNR) have a direct influence on the quality of the denoised speech. For instance, some common background noise types can be very similar to the desired speech such as cafe or food court noise. In these cases, estimating the desired speech from the corrupted signal is challenging and sometimes impossible in low SNR situations because the noise occupy the same frequency bands as the desired speech. This process of eliminating background noise from noisy speech signal is constructive for applications such as automatic speech recognition (ASR) systems, hearing aids and teleconferencing systems.

Previously, traditional approaches were adopted for speech enhancement such as spectral subtraction \cite{sep1,sep2} and binary masking techniques \cite{bm1,bm2}. Moreover, statistical approaches based on Wiener filters and Bayesian estimators were applied to speech enhancement \cite{wiener2,bayesian}. However, most of these approaches require a prior estimation of the SNR based on an initial silent period and can only operate well on limited non-speech like noise types in high SNR situations. This is attributed to the lack of a precise signal model describing the distinction between the speech and noise signals. 

To overcome such limitations, data driven approaches based on deep neural networks (DNNs) are widely used in literature to learn deep underlying features of either the desired speech or the intrusive background noise from the given data without a signal model. For instance, denoising autoencoders (AE) were used in \cite{dae1,dae2} to estimate a clean track from a noisy input based on the L1-loss. Long short-term memory (LSTM) networks have also been utilized to incorporate temporal speech structure in the denoising process \cite{lstm1,lstm2}. Also, an adaptation of the autoregressive generative WavNet was used in \cite{wavnet} where a denoised sample is generated based on the previous input and output samples.

In 2014, generative adversarial networks (GANs) were introduced as the state-of-the-art for deep generative models \cite{gan}. In GANs, a generator is trained adversarially with a discriminator to generate images belonging to the same joint distribution of the training data. Afterwards, variants of GANs such as conditional generative adversarial networks (cGANs) were introduced for image-to-image translation tasks \cite{cgan,cganMed,cganRad}. The pix2pix model introduced in \cite{pix2pix} is one of the first attempts to map natural images from an input source domain to a certain target domain. Henceforth, cGANs were used for speech enhancement either by utilizing the raw 1D speech tracks or the 2D log-Mel time-frequency (TF) magnitude representation. For instance, speech enhancement GAN (SEGAN) is a 1D adaptation of the pix2pix model operating on 1D raw speech tracks \cite{segan}. This model was further adapted to operate on 2D TF-magnitude representations via the frequency SEGAN (FSEGAN) framework \cite{fsegan}. Due to the TF-magnitude representation being used as an implicit feature extractor, an improved speech denoising was reported. However, both models suffer from multiple limitations. They rely mainly on pixel-wise losses, which have been reported to produce inconsistencies and output artifacts \cite{pix2pix}. Additionally, both models were utilized to denoise speech tracks of fixed durations and under relatively mild noise conditions with an average SNR of 10 dB.

In this work, a new adversarial approach, inspired by \cite{medgan}, is proposed for denoising speech tracks by operating on 2D TF-magnitude representations of noisy speech inputs. The proposed framework incorporates a cascaded architecture in addition to a non-adversarial feature-based loss which penalizes the discrepancies in the feature space between the outputs and the targets. This enhances the robustness of speech denoising with respect to harsh SNR conditions and speech-like background noise types. 

Additionally, we propose a new dynamic time resolution technique to embed variable track lengths in a fixed TF representation by adapting the time overlap according to the track length. To illustrate the performance of the proposed approach, quantitative comparison is carried out against SEGAN \cite{segan}, FSEGAN \cite{fsegan} and two traditional model-based variants of Wiener filters and Bayesian estimators \cite{wiener2,bayesian} under different noise types and SNR levels. Furthermore, the word error rate (WER) of a pre-trained automatic speech recognition (ASR) model is evaluated.
\begin{figure}
\vspace{-4mm}
	\centering
	\centerline{\hspace*{10mm}
		\captionsetup[subfigure]{oneside,margin={-0.8cm,0cm}}
		\subfloat[Track duration $=1.4\,s$.]
		{\resizebox{.68\columnwidth}{!}{\input{fig/track_1s.tex}}\label{fig:t1s}}
		\hspace{-14mm}\captionsetup[subfigure]{oneside,margin={-0.8cm,0cm}}
		\subfloat[Track duration $=4\,s$.]
		{\resizebox{.68\columnwidth}{!}{\input{fig/track_4s.tex}}\label{fig:t4s}}
	}
	\captionsetup{oneside,margin={0cm,0cm}}
	\caption{Examples of variable duration tracks embedded in a fixed $256\times256$ TF-magnitude representation. \label{fig:tracks}}
	\vspace{-4mm}
\end{figure}
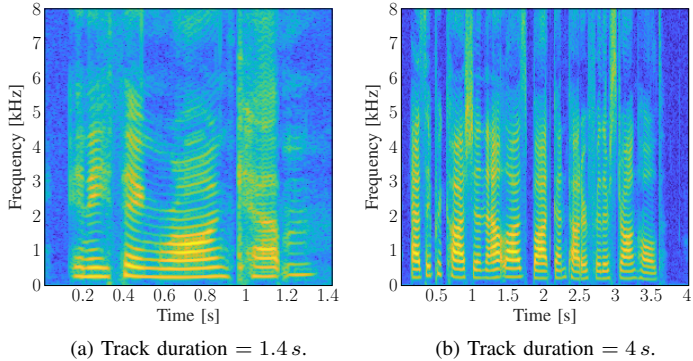
\vspace{-1mm}
\section{Dynamic Time Resolution} \label{sec:dtr}
Previously proposed architectures are designed to work on speech tracks of fixed durations. This is due to the architectural limitation of having to operate on inputs of fixed pixel dimensionality or number of samples for FSEGAN and SEGAN, respectively. In order to accommodate this constraint, the input track length was fixed to \unit[1]{s}. Accordingly, a track of arbitrary length should be first divided into \unit[1]{s} intervals and then the denoising is applied sequentially on each interval. 

In our proposed framework, the input 2D TF-magnitude representation is fixed to $256 \times 256$ pixels. However, the time resolution per pixel is variable according to the length of the 1D track as shown in Fig.~\ref{fig:tracks}. The TF-magnitude representation is computed based on short time Fourier transform (STFT) where a window function followed by FFT is applied to overlapping segments of the 1D track. In our case, we will consider tracks of \unit[16]{kHz} sampling frequency. A hamming window of $S=512$ samples is used to get a one-sided spectrum of $N_{F}=256$ frequency bins. To fix the time dimension to $N_{T} = 256$ time bins, the overlapping parameter $O$ of the 1D segments is adjusted based on the input track length $L$ according to the following relation: 
\begin{equation}
O = S-\ceil[\bigg]{\frac{L}{N_{T}}} \label{eq:overlap}
\end{equation}
Finally, the track length $L$ is modified either by omitting samples or padding a silent signal based on the following constraint:
\begin{equation}
L = N_{T}(S-O)+O \label{eq:tLength}
\end{equation}

\begin{figure}
	\centering
	\includegraphics[width=0.49\textwidth]{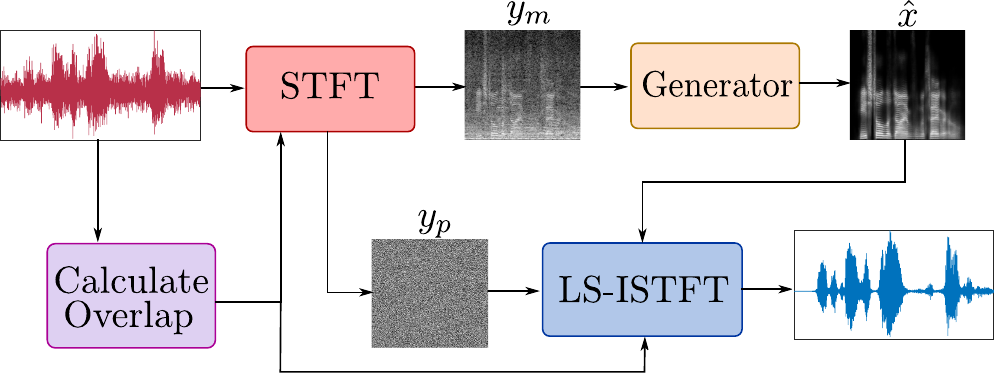}
	\caption{General block diagram of the proposed system. The log-magnitude is passed to the network and the output of the network is used with the input noisy phase for reconstruction.\label{fig:genArch}}
	\vspace{-3mm}
\end{figure}
After applying the denoising to the input TF-magnitude representation, getting back to the time domain is mandatory. For this we choose to use the least square inverse short time Fourier transform (LS-ISTFT) proposed in \cite{lsistft}. Based on this implementation, an acceptable signal-to-distortion ratio (SDR) reconstruction can be achieved with an overlap of at least 25\%. By substituting this overlap ratio in Eq.~\ref{eq:overlap}, the longest track length that can be embedded in a $256\times256$ TF-magnitude representation should not exceed \unit[6.1]{s}. Otherwise the track will be split into multiple suitable durations. The LS-ISTFT requires both magnitude and phase of the TF representation for reconstruction. However, the phonetic information of speech is mostly available in the magnitude. Therefore, only this magnitude $y_m$ is passed as input to the denoising network. For reconstruction, the noisy phase $y_p$ of the TF representation is used together with the denoised magnitude as shown in Fig.~\ref{fig:genArch}.

\section{Method}

\begin{figure}
	\centering
	\includegraphics[width=0.49\textwidth]{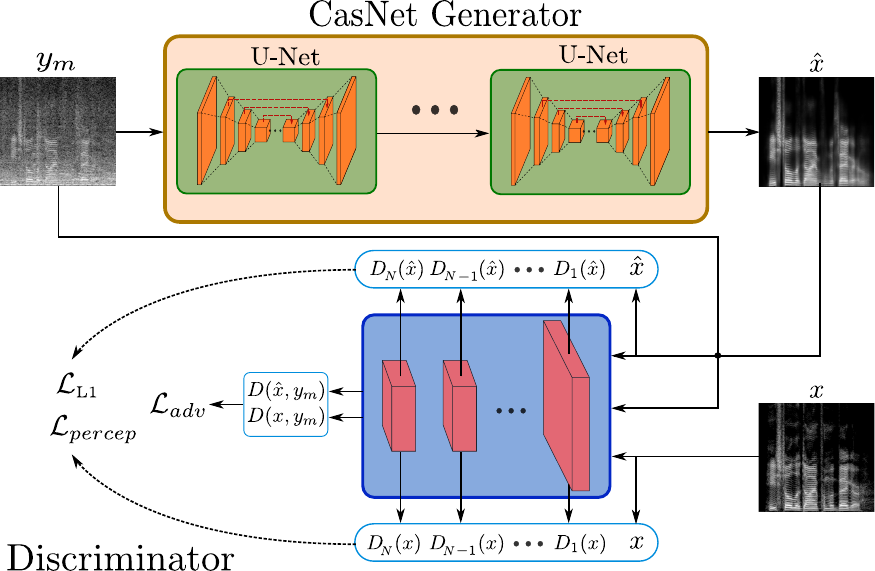}
	\caption{An overview of the proposed adversarial architecture for speech TF-magnitude denoising with relevant losses.\label{fig:netArch}}
	\vspace{-4mm}
\end{figure}

In this section, the proposed adversarial approach for speech enhancement will be described. First, a brief explanation of traditional cGANs will be outlined, followed by the proposed framework titled acoustic-enhancement GAN (AeGAN). However, in this initial work the AeGAN will be applied on a speech denoising task. An overview of the proposed approach is presented in Fig.~\ref{fig:netArch}.

\subsection{Conditional Generative Adversarial Networks}
In general, adversarial frameworks are a game-theoretical approach which pits multiple networks in direct competition with each other. More specifically, a cGAN framework consists of two deep convolutional neural networks (DCNNs), a generator $G$ and a discriminator $D$ \cite{pix2pix}. The generator receives as input the magnitude of the 2D TF representation of the noisy speech. It attempts to eliminate the intrusive background noise by outputting the denoised TF-magnitude $\hat{x} = G(y_m)$. The main goal of the generator is to render $\hat{x}$ to be indistinguishable from the target ground-truth clean speech TF-magnitude $x$. Parallel to this process, the discriminator network is trained to directly oppose the generator. $D$ acts as a binary classifier receiving $y_m$ and either $x$ or $\hat{x}$ as inputs and classifying which of them is synthetically generated and which is real. In other words, $G$ attempts to produce a realistically enhanced TF-magnitude to fool $D$, while conversely $D$ constantly improves it's performance to better detect the generator's output as fake. This adversarial training setting drives both networks to improve their respective performance until Nash's equilibrium is reached. This training procedure is expressed via the following min-max optimization task over the adversarial loss function $\mathcal{L}_{\small\textrm{adv}}$:
\begin{equation}
\begin{split}
\min_{G} \max_{D} \mathcal{L}_{\small\textrm{adv}} = \min_{G} \max_{D} & \; \mathbb{E}_{x,y_m} \left[\textrm{log} D(x,y_m) \right] + \\
& \;\mathbb{E}_{\hat{x},y_m} \left[\textrm{log} \left( 1 - D\left(\hat{x},y_m\right) \right) \right]
\end{split}
\end{equation}

To further improve the output of the generator and avoid visual artifacts, an additional L1 loss is utilized to enforce pixel-wise consistency between the generator output $\hat{x}$ and the ground-truth target \cite{pix2pix}. The L1 loss is given by
\begin{equation}
\mathcal{L}_{\small\textrm{L1}} = \mathbb{E}_{x,\hat{x}} \left[\lVert{x - \hat{x}}\rVert_1\right]
\end{equation}

\subsection{Feature-Based Loss}

The magnitude component of the speech TF representation has rich patterns directly reflecting human speech phonetics. A straightforward minimization of the pixel-wise discrepancy via L1 loss will result in a blurry TF-magnitude reconstruction which in turns will deteriorate the speech phonetics. 

To overcome this issue, we propose the utilization of the feature-based loss inspired by \cite{medgan} to regularize the generator network to produce globally consistent results by focusing on wider feature representations rather than individual pixels. This is achieved by utilizing the discriminator $D$ as a trainable feature extractor to extract low and high-level feature representations. The feature-based loss is then calculated as the weighted average of the mean absolute error (MAE) of the extracted feature maps:

\begin{equation}
\mathcal{L}_{\small\textrm{Percep}} = \sum_{i = 1}^{N}  \lambda_{n} \lVert{D_n\left(x\right) - D_n\left(\hat{x}\right)}\rVert_1
\end{equation}
where $D_n$ is the feature map extracted from the $n^{th}$ layer of the discriminator. $N$ and $\lambda_n$ are the total number of layers and the individual weights given to each layer, respectively.
\subsection{Architectural Details}

In our proposed AeGAN framework, a CasNet generator and a patch discriminator architecture are utilized \cite{medgan}. CasNet concatenates three U-blocks in an end-to-end manner, whereas each U-block consists of a encoder-decoder architecture joint together via skip connections. These connections avoid the excessive loss of information due to the bottleneck layer. The output TF-magnitude representations are progressively refined as they propagate through the multiple encoder-decoder pairs. The architecture of each U-block is identical to that proposed in \cite{pix2pix}.  Regarding the patch discriminator, it divides the input TF-magnitude representations into smaller patches before proceeding with classifying each patch as real or fake. For the final classification score, all patch scores are averaged out. However, unlike the $70 \times 70$ pixel patches recommended in \cite{pix2pix}, a patch size of $16 \times 16$ was found to produce better output results in our case.

\section{Experiments}

\begin{figure*}
	\centering
	{\resizebox{0.94\textwidth}{!}{\input{fig/qual.tex}}}
	\captionsetup{justification=centering}
	\caption{Qualitative comparison on the TF-magnitude of different learning-based speech denoising techniques.
		\newline (\myelipsegreen) illustrates the frequency bands of different noise types. (\myarrowred) and (\myarrowblue) shows the advantages of our model under cafe noise. \label{fig:qual}
		\vspace{-1mm}}
\end{figure*}
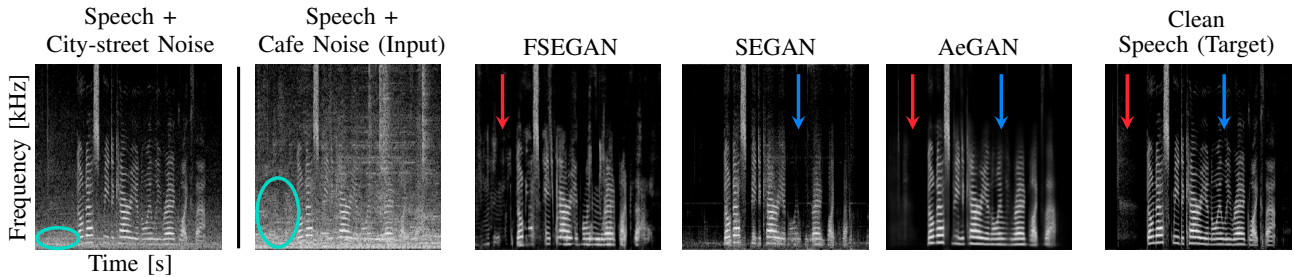

\begin{table*}[!ht]
	\vspace{-1mm}
	\caption{Quantitative comparison of speech denoising techniques on test dataset with speech-like noise types.\label{tab:test}}
	\vspace{-1mm}
	\centering
	\setlength\arrayrulewidth{0.05pt}
	\normalsize
	\bgroup
	\resizebox{\linewidth}{!}{%
		\begin{tabular}{r|ccc|ccc|ccc|ccc|ccc|ccc|c}
			\hline\hline
			\multirow{3}{*}{Model} & \multicolumn{3}{c|}{Noisy input} & \multicolumn{3}{c|}{Wiener filter \cite{wiener2}} & \multicolumn{3}{c|}{Bayesian est. \cite{bayesian}} & \multicolumn{3}{c|}{FSEGAN \cite{fsegan}} & \multicolumn{3}{c|}{SEGAN \cite{segan}} & \multicolumn{3}{c|}{AeGAN}& \multirow{3}{*}{Target}\\ \cmidrule(lr){2-4}\cmidrule(lr){5-7}\cmidrule(lr){8-10}\cmidrule(lr){11-13}\cmidrule(lr){14-16}\cmidrule(lr){17-19} & \multicolumn{3}{c|}{SNR (dB)} & \multicolumn{3}{c|}{SNR (dB)} & \multicolumn{3}{c|}{SNR (dB)} & \multicolumn{3}{c|}{SNR (dB)} & \multicolumn{3}{c|}{SNR (dB)} & \multicolumn{3}{c|}{SNR (dB)}\\ & 0 & 5 & 10 & 0 & 5 & 10 & 0 & 5 & 10 & 0 & 5 & 10 & 0 & 5 & 10 & 0 & 5 & 10\\
			\hline
			PESQ & 1.43 & 1.69 & 2.03 & 1.46 & 1.79 & 2.20 & 1.45 & 1.79 & 2.21 & 1.57 & 1.99 & 2.47 & 1.79 & 2.18 & 2.59 & \textbf{2.17} & \textbf{2.64} & \textbf{3.04} & \textbf{4.5}\\
			CSIG & 2.23 & 2.75 & 3.29 & 1.79 & 2.43 & 3.04 & 1.81 & 2.41 & 2.98 & 2.50 & 3.12 & 3.69 & 2.83 & 3.34 & 3.78 & \textbf{3.36} & \textbf{3.86} & \textbf{4.28} & \textbf{5.0}\\
			CBAK & 1.64 & 2.03 & 2.49 & 1.58 & 2.06 & 2.58 & 1.66 & 2.10 & 2.59 & 2.11 & 2.54 & 2.97 & 2.26 & 2.65 & 3.01 & \textbf{2.59} & \textbf{3.00} & \textbf{3.35} & \textbf{5.0}\\
			COVL & 1.72 & 2.14 & 2.61 & 1.45 & 1.98 & 2.53 & 1.47 & 1.98 & 2.51 & 1.97 & 2.52 & 3.06 & 2.23 & 2.71 & 3.16 & \textbf{2.74} & \textbf{3.24} & \textbf{3.66} & \textbf{5.0}\\
			STOI & 0.65 & 0.77 & 0.86 & 0.63 & 0.76 & 0.86 & 0.60 & 0.73 & 0.83 & 0.72 & 0.82 & 0.89 & 0.79 & 0.86 & 0.91 & \textbf{0.82} & \textbf{0.89} & \textbf{0.93} & \textbf{1.0}\\
			ASR WER (\%) & 90.0 & 70.4 & 46.3 & 89.6 & 70.0 & 49.6 & 87.7 & 71.3 & 49.0 & 85.9 & 66.4 & 44.7 & 73.8 & 53.7 & 35.8 & \textbf{64.6} & \textbf{42.9} & \textbf{29.7} & \textbf{20.0}\\
			\hline\hline
		\end{tabular}
	}
	\egroup
\end{table*}
\begin{table*}[!ht]
	\vspace{-1mm}
	\caption{Quantitative results for generalization on city-street noise.\label{tab:gen}}
	\vspace{-1mm}
	\centering
	\setlength\arrayrulewidth{0.05pt}
	\normalsize
	\bgroup
	\resizebox{\linewidth}{!}{%
		\begin{tabular}{r|ccc|ccc|ccc|ccc|ccc|ccc|c}
			\hline\hline
			\multirow{3}{*}{Model} & \multicolumn{3}{c|}{Noisy Input} & \multicolumn{3}{c|}{Wiener filter \cite{wiener2}} & \multicolumn{3}{c|}{Bayesian est. \cite{bayesian}} & \multicolumn{3}{c|}{FSEGAN \cite{fsegan}} & \multicolumn{3}{c|}{SEGAN \cite{segan}} & \multicolumn{3}{c|}{AeGAN} & \multirow{3}{*}{Target}\\ \cmidrule(lr){2-4}\cmidrule(lr){5-7}\cmidrule(lr){8-10}\cmidrule(lr){11-13}\cmidrule(lr){14-16}\cmidrule(lr){17-19} & \multicolumn{3}{c|}{SNR (dB)} & \multicolumn{3}{c|}{SNR (dB)} & \multicolumn{3}{c|}{SNR (dB)} & \multicolumn{3}{c|}{SNR (dB)} & \multicolumn{3}{c|}{SNR (dB)} & \multicolumn{3}{c|}{SNR (dB)}\\ & 0 & 5 & 10 & 0 & 5 & 10 & 0 & 5 & 10 & 0 & 5 & 10 & 0 & 5 & 10 & 0 & 5 & 10\\
			\hline
			PESQ & 1.46 & 1.73 & 2.13 & 1.72 & 2.11 & 2.58 & 1.81 & 2.22 & 2.69 & 1.72 & 2.19 & 2.70 & 1.81 & 2.22 & 2.64 & \textbf{2.47} & \textbf{2.90} & \textbf{3.27} & \textbf{4.5}\\
			CSIG & 2.40 & 2.93 & 3.49 & 2.52 & 3.11 & 3.63 & 2.54 & 3.11 & 3.60 & 2.73 & 3.40 & 3.99 & 3.00 & 3.51 & 3.93 & \textbf{3.81} & \textbf{4.24} & \textbf{4.59} & \textbf{5.0}\\
			CBAK & 1.55 & 1.97 & 2.49 & 1.83 & 2.34 & 2.88 & 1.99 & 2.48 & 2.98 & 2.25 & 2.71 & 3.16 & 2.29 & 2.71 & 3.09 & \textbf{2.84} & \textbf{3.22} & \textbf{3.57} & \textbf{5.0}\\
			COVL & 1.79 & 2.23 & 2.75 & 1.96 & 2.50 & 3.03 & 2.03 & 2.56 & 3.08 & 2.16 & 2.76 & 3.33 & 2.32 & 2.81 & 3.26 & \textbf{3.12} & \textbf{3.56} & \textbf{3.94} & \textbf{5.0}\\
			STOI & 0.72 & 0.80 & 0.87 & 0.72 & 0.80 & 0.88 & 0.70 & 0.78 & 0.85 & 0.75 & 0.84 & 0.89 & 0.81 & 0.88 & 0.92 & \textbf{0.86} & \textbf{0.90} & \textbf{0.94} & \textbf{1.0}\\
			ASR WER (\%) & 85.6 & 64.7 & 45.4 & 78.0 & 58.2 & 38.9 & 75.5 & 56.8 & 40.5 & 81.2 & 62.3 & 43.0 & 70.2 & 51.3 & 34.7 & \textbf{50.1} & \textbf{39.4} & \textbf{27.1} & \textbf{20.0}\\
			\hline \hline
		\end{tabular}
	}
	\egroup
	\vspace{-5mm}
\end{table*}
The proposed speech denoising framework is evaluated on the TIMIT dataset \cite{timit}. This dataset consists of 10 phonetically rich sentences spoken by 630 speakers with 8 different American English dialects. All tracks are sampled at \unit[16]{kHz} and the track durations are between \unit[0.9]{s} to \unit[7]{s}. The majority of the tracks satisfies the aforementioned track length constraint in Sec.~\ref{sec:dtr}. Only 15 tracks were found to exceed the \unit[6.1]{s} limit and were excluded from the dataset for simplicity. 

In the training procedure, three different noise types were utilized (cafe, food court and home kitchen) from the QUT-TIMIT proposed in \cite{qutnoise}. The background noise was added to the clean speech in order to create a paired training set. Additionally, different total SNR levels were used for each noise type (0, 5 and 10 dB). Thus, the total training dataset consists of 36,000 paired tracks from 462 speakers of 6 different dialects. For validation, two different experiments were conducted. In the first experiment, the trained network was validated on a test set of 5000 tracks utilizing the same training noise types albeit from different 168 individuals using the whole 8 available dialects. In the second experiment, the generalization capability of the network was investigated by validating on a dataset of 500 tracks from the test set corrupted by a new noise type, the city street noise. Both experiments were conducted using the same SNR values used in training. 

To compare the performance of the proposed approach, quantitative comparisons were conduced against the FSEGAN and SEGAN \cite{fsegan,segan}. Additionally, traditional model-based approaches, Wiener filter \cite{wiener2} and an optimized weighted-Euclidean Bayesian estimator \cite{bayesian}, were utilized in the comparative study based on their open-source implementations\footnote{\url{https://www.crcpress.com/downloads/K14513/K14513_CD_Files.zip}}. All trainable models were trained using the same hyperparameters for 50 epochs to ensure a fair comparison. Multiple metrics were used for the comparison in order to give a wider scope of interpretation for the results. The utilized metrics are the perceptual evaluation of speech quality (PSEQ) \cite{pesq}, the mean opinion score (MOS) prediction of the signal distortion (CSIG), the MOS prediction of background noise (CBAK) and the overall MOS prediction score (COVL) \cite{mos}. To give an indication of human speech intelligibility, the short-time objective intelligibility measure (STOI) was utilized \cite{stoi}. Additionally, the WER was evaluated using the Deep Speech pre-trained ASR model \cite{deepspeech}.
\vspace{-0.5mm}
\section{Results and Discussion}

First a qualitative comparison of the TF-magnitude representation of different methods is illustrated. As shown in Fig.~\ref{fig:qual}, the AeGAN is superior in cancelling the low power components of the background noise in comparison to FSEGAN as annotated by (\hspace{-0.71mm}\myarrowred\hspace{-0.71mm}). In contrast to the AeGAN, the SEGAN model shows a clear elimination of some speech intervals as annotated by (\hspace{-0.3mm}\myarrowblue\hspace{-0.3mm}). 

Regarding the quantitative analysis, we present the metric scores of the noisy input tracks as a comparison baseline. Also the metric scores of the ground-truth target clean tracks are presented as an indicator of the maximal achievable performance. All scores are averaged over the different noise types. In Table~\ref{tab:test}, the results of the first experiment is presented. In this experiment, the test tracks were based on speech-like noise types (cafe and food-court noise). Hence, the noise distribution is difficult to distinguish from the target speech segments. The model-based approaches resulted in minor or no speech improvements compared to the baseline noisy input. We hypothesize that this is due to the fact that the distribution of the speech-like noise occupies the same frequency bands as the speech signals. Thus, the model-based approaches fail to distinguish the speech from the noisy background in case of speech-like corruption. Regarding the learning-based approaches, the proposed AeGAN framework outperforms both the FSEGAN and SEGAN models. For instance, AeGAN results in a WER of 29.7\% for SNR \unit[10]{dB} compared to 35.8\% and 44.7\% for SEGAN and FSEGAN, respectively. 

To illustrate the generalization capability of the proposed framework, an additional comparative study is presented in Table~\ref{tab:gen} based on validating the trained models on a new noise type (city-street noise). This noise can be considered as a less challenging noise compared to the aforementioned speech-like noises because it occupies a narrower frequency band as annotated by (\hspace{-0.3mm}\myelipsegreen\hspace{-0.3mm}) in Fig.~\ref{fig:qual}. Accordingly, the model-based approaches resulted in a more noticeable improvement compared to the noisy input baseline. More specifically, the more recent Bayesian estimator outperformed the traditional Wiener filter across the objective metrics. FSEGAN resulted in an enhanced performance in the objective metrics with slight deterioration in the PESQ and WER compared to the model-based approaches. Finally, the SEGAN and AeGAN are quantitatively superior across all utilized metrics with AeGAN enhancing the WER by 20.1\% compared to SEGAN in the \unit[0]{dB} case. This illustrates that the learning-based approaches result in a significant improvement in speech denoising performance with robust generalization to never seen noise types, especially SEGAN and the proposed AeGAN. 

Conventionally, deep-learning approaches face a significant challenge in collecting a large enough number of labeled training samples, i.e. paired (clean and noisy) samples. However, in the case of speech denoising this is easily bypassed by the availability of accessible audio and noise datasets that can be superimposed with the required SNR. 

It must also be pointed that in literature the FSEGAN authors claim a better performance in WER over the SEGAN model.  However, this has not been observed in the above results. We hypothesize that this is the result of FSEGAN now having to deal with variable time resolution input TF-magnitude representations, due to the utilized dynamic time resolution, which posses a challenge compared to the SEGAN.

However, this work is not without limitation. In the future, we plan to extend the current comparative studies to include more recent model-based approaches for speech denoising such as \cite{mixture_model,bayesian_2018}. In addition to applying some subjective evaluation tests.We also plan to extend the AeGAN framework to accommodate different non-speech audio signals (e.g. music denoising) and other enhancement tasks such as dereverberation and interference cancellation. 
\vspace{-2mm}
\section{Conclusion}
\vspace{-1mm}
In this work, an adversarial speech denoising technique is introduced to operate on speech TF-magnitude representations. The proposed approach involves an additional feature-based loss and a CasNet generator architecture to enhance detailed local features of speech in the TF domain. Moreover, to improve the inference efficiency, time-domain tracks with variable durations are embedded in a fixed TF-magnitude representation by changing the corresponding time resolution. 

Challenging speech-like noise types, e.g. cafe and food court noise, were involved in training under low SNR conditions. To evaluate the generalization capability of our model, two experiments were conducted on different speakers and noise types. The proposed approach exhibits a significantly enhanced performance in comparison to the previously introduced GAN-based and traditional model-based approaches. 
\vspace{-1.5mm}
\bibliographystyle{IEEEtran}

\end{document}

%% file: fig/track_1s.tex
%
%
\begin{tikzpicture}
\newcommand\mtlarge{\fontsize{17pt}{18pt}\selectfont}
\begin{axis}[%
width=3.7in,
height=3.566in,
at={(0.676in,0.481in)},
scale only axis,
point meta min=-100,
point meta max=0,
axis on top,
xmin=0.01321875,
xmax=1.42121875,
xlabel style={font=\color{white!15!black}},
ymin=-0.0156555772994129,
ymax=8,
ylabel style={font=\color{white!15!black}},
xlabel={\mtlarge Time [s]},
ylabel={\mtlarge Frequency [kHz]},
ytick style={draw=none},
xtick style={draw=none},
yticklabel style = {font=\mtlarge},
xticklabel style = {font=\mtlarge},
axis background/.style={fill=white},
legend style={legend cell align=left, align=left, draw=white!15!black},
colormap={mymap}{[1pt] rgb(0pt)=(0.2422,0.1504,0.6603); rgb(1pt)=(0.25039,0.164995,0.707614); rgb(2pt)=(0.257771,0.181781,0.751138); rgb(3pt)=(0.264729,0.197757,0.795214); rgb(4pt)=(0.270648,0.214676,0.836371); rgb(5pt)=(0.275114,0.234238,0.870986); rgb(6pt)=(0.2783,0.255871,0.899071); rgb(7pt)=(0.280333,0.278233,0.9221); rgb(8pt)=(0.281338,0.300595,0.941376); rgb(9pt)=(0.281014,0.322757,0.957886); rgb(10pt)=(0.279467,0.344671,0.971676); rgb(11pt)=(0.275971,0.366681,0.982905); rgb(12pt)=(0.269914,0.3892,0.9906); rgb(13pt)=(0.260243,0.412329,0.995157); rgb(14pt)=(0.244033,0.435833,0.998833); rgb(15pt)=(0.220643,0.460257,0.997286); rgb(16pt)=(0.196333,0.484719,0.989152); rgb(17pt)=(0.183405,0.507371,0.979795); rgb(18pt)=(0.178643,0.528857,0.968157); rgb(19pt)=(0.176438,0.549905,0.952019); rgb(20pt)=(0.168743,0.570262,0.935871); rgb(21pt)=(0.154,0.5902,0.9218); rgb(22pt)=(0.146029,0.609119,0.907857); rgb(23pt)=(0.138024,0.627629,0.89729); rgb(24pt)=(0.124814,0.645929,0.888343); rgb(25pt)=(0.111252,0.6635,0.876314); rgb(26pt)=(0.0952095,0.679829,0.859781); rgb(27pt)=(0.0688714,0.694771,0.839357); rgb(28pt)=(0.0296667,0.708167,0.816333); rgb(29pt)=(0.00357143,0.720267,0.7917); rgb(30pt)=(0.00665714,0.731214,0.766014); rgb(31pt)=(0.0433286,0.741095,0.73941); rgb(32pt)=(0.0963952,0.75,0.712038); rgb(33pt)=(0.140771,0.7584,0.684157); rgb(34pt)=(0.1717,0.766962,0.655443); rgb(35pt)=(0.193767,0.775767,0.6251); rgb(36pt)=(0.216086,0.7843,0.5923); rgb(37pt)=(0.246957,0.791795,0.556743); rgb(38pt)=(0.290614,0.79729,0.518829); rgb(39pt)=(0.340643,0.8008,0.478857); rgb(40pt)=(0.3909,0.802871,0.435448); rgb(41pt)=(0.445629,0.802419,0.390919); rgb(42pt)=(0.5044,0.7993,0.348); rgb(43pt)=(0.561562,0.794233,0.304481); rgb(44pt)=(0.617395,0.787619,0.261238); rgb(45pt)=(0.671986,0.779271,0.2227); rgb(46pt)=(0.7242,0.769843,0.191029); rgb(47pt)=(0.773833,0.759805,0.16461); rgb(48pt)=(0.820314,0.749814,0.153529); rgb(49pt)=(0.863433,0.7406,0.159633); rgb(50pt)=(0.903543,0.733029,0.177414); rgb(51pt)=(0.939257,0.728786,0.209957); rgb(52pt)=(0.972757,0.729771,0.239443); rgb(53pt)=(0.995648,0.743371,0.237148); rgb(54pt)=(0.996986,0.765857,0.219943); rgb(55pt)=(0.995205,0.789252,0.202762); rgb(56pt)=(0.9892,0.813567,0.188533); rgb(57pt)=(0.978629,0.838629,0.176557); rgb(58pt)=(0.967648,0.8639,0.16429); rgb(59pt)=(0.96101,0.889019,0.153676); rgb(60pt)=(0.959671,0.913457,0.142257); rgb(61pt)=(0.962795,0.937338,0.12651); rgb(62pt)=(0.969114,0.960629,0.106362); rgb(63pt)=(0.9769,0.9839,0.0805)},
]
\addplot [forget plot] graphics [xmin=0.01321875, xmax=1.42121875, ymin=-0.0156555772994129, ymax=8] {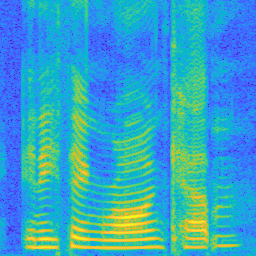};
\end{axis}

\begin{axis}[%
width=5.833in,
height=4.375in,
at={(0in,0in)},
scale only axis,
point meta min=0,
point meta max=1,
xmin=0,
xmax=1,
ymin=0,
ymax=1,
axis line style={draw=none},
ticks=none,
axis x line*=bottom,
axis y line*=left,
legend style={legend cell align=left, align=left, draw=white!15!black}
]
\end{axis}
\end{tikzpicture}%

%% file: fig/track_4s.tex
%
%
\begin{tikzpicture}
\newcommand\mtlarge{\fontsize{17pt}{18pt}\selectfont}
\begin{axis}[%
width=3.7in,
height=3.566in,
at={(0.671in,0.481in)},
scale only axis,
point meta min=-100,
point meta max=0,
axis on top,
xmin=0.008125,
xmax=4.024125,
xlabel style={font=\color{white!15!black}},
ymin=-0.0156555772994129,
ymax=8,
ylabel style={font=\color{white!15!black}},
xlabel={\mtlarge Time [s]},
ylabel={\mtlarge Frequency [kHz]},
ytick style={draw=none},
xtick style={draw=none},
yticklabel style = {font=\mtlarge},
xticklabel style = {font=\mtlarge},
axis background/.style={fill=white},
legend style={legend cell align=left, align=left, draw=white!15!black},
colormap={mymap}{[1pt] rgb(0pt)=(0.2422,0.1504,0.6603); rgb(1pt)=(0.25039,0.164995,0.707614); rgb(2pt)=(0.257771,0.181781,0.751138); rgb(3pt)=(0.264729,0.197757,0.795214); rgb(4pt)=(0.270648,0.214676,0.836371); rgb(5pt)=(0.275114,0.234238,0.870986); rgb(6pt)=(0.2783,0.255871,0.899071); rgb(7pt)=(0.280333,0.278233,0.9221); rgb(8pt)=(0.281338,0.300595,0.941376); rgb(9pt)=(0.281014,0.322757,0.957886); rgb(10pt)=(0.279467,0.344671,0.971676); rgb(11pt)=(0.275971,0.366681,0.982905); rgb(12pt)=(0.269914,0.3892,0.9906); rgb(13pt)=(0.260243,0.412329,0.995157); rgb(14pt)=(0.244033,0.435833,0.998833); rgb(15pt)=(0.220643,0.460257,0.997286); rgb(16pt)=(0.196333,0.484719,0.989152); rgb(17pt)=(0.183405,0.507371,0.979795); rgb(18pt)=(0.178643,0.528857,0.968157); rgb(19pt)=(0.176438,0.549905,0.952019); rgb(20pt)=(0.168743,0.570262,0.935871); rgb(21pt)=(0.154,0.5902,0.9218); rgb(22pt)=(0.146029,0.609119,0.907857); rgb(23pt)=(0.138024,0.627629,0.89729); rgb(24pt)=(0.124814,0.645929,0.888343); rgb(25pt)=(0.111252,0.6635,0.876314); rgb(26pt)=(0.0952095,0.679829,0.859781); rgb(27pt)=(0.0688714,0.694771,0.839357); rgb(28pt)=(0.0296667,0.708167,0.816333); rgb(29pt)=(0.00357143,0.720267,0.7917); rgb(30pt)=(0.00665714,0.731214,0.766014); rgb(31pt)=(0.0433286,0.741095,0.73941); rgb(32pt)=(0.0963952,0.75,0.712038); rgb(33pt)=(0.140771,0.7584,0.684157); rgb(34pt)=(0.1717,0.766962,0.655443); rgb(35pt)=(0.193767,0.775767,0.6251); rgb(36pt)=(0.216086,0.7843,0.5923); rgb(37pt)=(0.246957,0.791795,0.556743); rgb(38pt)=(0.290614,0.79729,0.518829); rgb(39pt)=(0.340643,0.8008,0.478857); rgb(40pt)=(0.3909,0.802871,0.435448); rgb(41pt)=(0.445629,0.802419,0.390919); rgb(42pt)=(0.5044,0.7993,0.348); rgb(43pt)=(0.561562,0.794233,0.304481); rgb(44pt)=(0.617395,0.787619,0.261238); rgb(45pt)=(0.671986,0.779271,0.2227); rgb(46pt)=(0.7242,0.769843,0.191029); rgb(47pt)=(0.773833,0.759805,0.16461); rgb(48pt)=(0.820314,0.749814,0.153529); rgb(49pt)=(0.863433,0.7406,0.159633); rgb(50pt)=(0.903543,0.733029,0.177414); rgb(51pt)=(0.939257,0.728786,0.209957); rgb(52pt)=(0.972757,0.729771,0.239443); rgb(53pt)=(0.995648,0.743371,0.237148); rgb(54pt)=(0.996986,0.765857,0.219943); rgb(55pt)=(0.995205,0.789252,0.202762); rgb(56pt)=(0.9892,0.813567,0.188533); rgb(57pt)=(0.978629,0.838629,0.176557); rgb(58pt)=(0.967648,0.8639,0.16429); rgb(59pt)=(0.96101,0.889019,0.153676); rgb(60pt)=(0.959671,0.913457,0.142257); rgb(61pt)=(0.962795,0.937338,0.12651); rgb(62pt)=(0.969114,0.960629,0.106362); rgb(63pt)=(0.9769,0.9839,0.0805)},
]
\addplot [forget plot] graphics [xmin=0.008125, xmax=4.024125, ymin=-0.0156555772994129, ymax=8] {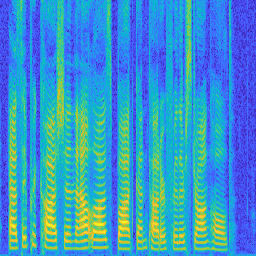};
\end{axis}

\begin{axis}[%
width=5.833in,
height=4.375in,
at={(0in,0in)},
scale only axis,
point meta min=0,
point meta max=1,
xmin=0,
xmax=1,
ymin=0,
ymax=1,
axis line style={draw=none},
ticks=none,
axis x line*=bottom,
axis y line*=left,
legend style={legend cell align=left, align=left, draw=white!15!black}
]
\end{axis}
\end{tikzpicture}%

%% file: fig/qual.tex
\begin{tikzpicture}
\pgfmathsetmacro{\PHI}{-4.5}
\pgfmathsetmacro{\Y}{3.6}
\pgfmathsetmacro{\Yshift}{3.85}
\pgfmathsetmacro{\Yshifta}{3.6}
\pgfmathsetmacro{\Yshiftb}{3.42}
\pgfmathsetmacro{\O}{3.2}
\pgfmathsetmacro{\P}{2.4}
\pgfmathsetmacro{\X}{0.5mm}
\node[inner sep=0pt] () at (0,2)
{\includegraphics[width=\textwidth]{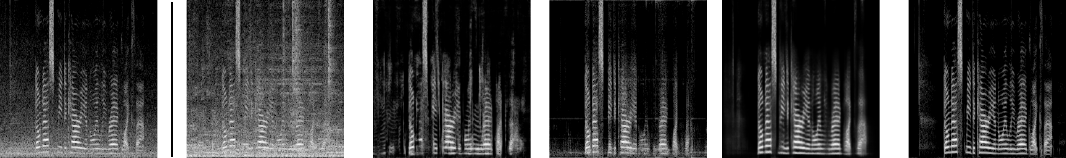}};
\node[inner sep=0pt] () at (-7.7, \Y+0.4) {Speech +};
\node[inner sep=0pt] () at (-7.7, \Y) {City-street Noise};
\node[inner sep=0pt] () at (-7.7, \Y-3.15) {Time [s]};
\node[inner sep=0pt,rotate=90] () at (-9.3, \Y-1.6) {Frequency [kHz]};
\node[inner sep=0pt] () at (-4.5, \Y+0.4) {Speech +};
\node[inner sep=0pt] () at (-4.5, \Y) {Cafe Noise (Input)};
\node[inner sep=0pt] () at (-1.34,\Y) {FSEGAN};
\node[inner sep=0pt] () at (1.62, \Y) {SEGAN};
\node[inner sep=0pt] () at (4.55, \Y) {AeGAN};
\node[inner sep=0pt] () at (7.7, \Y+0.4) {Clean};
\node[inner sep=0pt] () at (7.7, \Y) {Speech (Target)};
\draw [line width=\X,green2](-8.75, \O-2.37) ellipse (0.3cm and 0.15cm);
\draw [line width=\X,green2](-5.57, \O-2) ellipse (0.3cm and 0.5cm);
\draw [->,>=stealth,line width=\X,red2] (-2.325, \O) -- (-2.325, \P);
\draw [->,>=stealth,line width=\X,red2] (3.6, \O) -- (3.6, \P);
\draw [->,>=stealth,line width=\X,red2] (6.7, \O) -- (6.7, \P);
\draw [->,>=stealth,line width=\X,blue2] (1.95, \O) -- (1.95, \P); 
\draw [->,>=stealth,line width=\X,blue2] (4.88, \O) -- (4.88, \P);
\draw [->,>=stealth,line width=\X,blue2] (8.1, \O) -- (8.1, \P); 
\end{tikzpicture}%